# A systematic review of smartphone-based human activity recognition for health research

Marcin Straczkiewicz[1]*, Peter James[2,3], Jukka-Pekka Onnela[4]

[1] Department of Biostatistics, Harvard T.H. Chan School of Public Health, Boston, MA 02115, USA; mstraczkiewicz@hsph.harvard.edu

[2] Department of Population Medicine, Harvard Medical School and Harvard Pilgrim Health Care Institute, Boston, MA 02215, USA; pjames@hsph.harvard.edu

[3] Department of Environmental Health, Harvard T.H. Chan School of Public Health, Boston, MA 02115, USA; pjames@hsph.harvard.edu

[4] Department of Biostatistics, Harvard T.H. Chan School of Public Health, Boston, MA 02115, USA; onnela@hsph.harvard.edu

* Corresponding author

## Abstract

**Background:** Smartphones are now nearly ubiquitous; their numerous built-in sensors enable continuous measurement of activities of daily living, making them especially well-suited for health research. Researchers have proposed various human activity recognition (HAR) systems aimed at translating measurements from smartphones into various types of physical activity. In this review, we summarize the existing approaches to smartphone-based HAR.

**Methods:** We systematically searched Scopus, PubMed, and Web of Science for peer-reviewed articles published up to December 2020 on the use of smartphones for HAR. We extracted information on smartphone body location, sensors, and physical activity types studied and the data transformation techniques and classification schemes used for activity recognition.

**Results:** We identified 108 articles and described the various approaches used for data acquisition, data preprocessing, feature extraction, and activity classification, identifying the most common practices and their alternatives.

**Conclusions:** Smartphones are well-suited for HAR research in the health sciences. For population-level impact, future studies should focus on improving quality of collected data, address missing data, incorporate more diverse participants and activities, relax requirements about phone placement, provide more complete documentation on study participants, and share the source code of the implemented methods and algorithms.

## 1. Introduction

Progress in science has always been driven by data. More than 5 billion mobile devices were in use in 2020 [1], with multiple sensors (e.g., accelerometer and GPS) that can capture detailed, continuous, and objective measurements on various aspects of our lives, including physical activity. Such proliferation in worldwide smartphone adoption presents unprecedented opportunities for the collection of data to study human behavior and health. Along with sufficient storage, powerful processors, and wireless transmission, smartphones can collect a tremendous amount of data on large cohorts of individuals over extended time periods without additional hardware or instrumentation.

Smartphones are promising data collection instruments for objective and reproducible quantification of traditional and emerging risk factors for human populations. Behavioral risk factors, including but not limited to sedentary behavior, sleep, and physical activity, can all be monitored by smartphones in free-living environments, leveraging the personal or lived experiences of individuals. Importantly, unlike some wearable activity trackers [2], smartphones are not a niche product but instead have become globally available, increasingly adopted by users of all ages both in advanced and emerging economies [3,4]. Their adoption in health research is further supported by encouraging findings made with other portable devices, primarily wearable accelerometers, which have demonstrated robust associations between physical activity and health outcomes, including obesity, diabetes, various cardiovascular diseases, mental health, and mortality [5–9]. However, there are some important limitations to using wearables for studying population health: (1) their ownership is much lower than that of smartphones [10]; (2) most people stop using their wearables after 6 months of use [11]; and (3) raw data is usually not available from wearable devices. The last point often forces investigators to rely on proprietary device metrics, which lowers the already low rate of reproducibility of biomedical research in general [12], and makes uncertainty quantification in the measurements nearly impossible.

Human activity recognition (HAR) is a process aimed at classification of human actions in a given period of time based on discrete measurements (acceleration, rotation speed, geographical coordinates, etc.) made by personal digital devices. In recent years, this topic has been proliferating within the machine learning research community; at the time of writing, over 400 articles had been published on HAR methods using smartphones. This is a substantial increase from just a handful of articles published a few years earlier (Figure 1). As data collection using smartphones becomes easier, analysis of the collected data is increasingly identified as the main bottleneck in health research [13–15]. To tackle the analytical challenges of HAR, researchers have proposed various algorithms that differ substantially in terms of the type of data they use, how they manipulate the collected data, and the statistical approaches used for inference and/or classification. Published studies use existing methods and propose new methods for collection, processing, and classification of activities of daily living. Authors commonly discuss data filtering and feature selection techniques and compare the accuracy of various machine learning classifiers either on previously existing datasets or on datasets they have collected *de novo* for the purposes of the specific study. The results are typically summarized using classification accuracy within different groups of activities, such as ambulation, locomotion, and exercise.

To successfully incorporate developments in HAR into research in public health and medicine, there is a need to understand the approaches that have been developed and identify their potential limitations. Methods need to accommodate physiological (e.g., weight, height, age) and habitual (e.g., posture, gait, walking speed) differences of smartphone users, as well as differences in the built environment (e.g., buildings and green spaces) that provide the physical and social setting for human activities. Moreover, the data collection and statistical approaches typically used in HAR may be affected by location (where the user wears the phone on their body) and orientation of the device [16], which complicates the transformation of collected data into meaningful and interpretable outputs.

In this paper, we systematically review the emerging literature on the use of smartphones for HAR for health research in free-living settings. Given that the main challenge in this field is shifting from data collection to data analysis, we focus our analysis on the approaches used for data acquisition, data preprocessing, feature extraction, and activity classification. We provide insight into the complexity and multidimensionality of HAR utilizing smartphones, the types of data collected, and the methods used to translate digital measurements into human activities. We discuss the generalizability and reproducibility of approaches, i.e., the features that are essential and applicable to large and diverse cohorts of study participants. Lastly, we identify challenges that need to be tackled to accelerate wider utilization of smartphone-based HAR in public health studies.

(**Figure 1** About here)

## 2. Methods

Our systematic review was conducted by searching for articles published up to **December 31, 2020**, on PubMed, Scopus, and Web of Science databases. The databases were screened for titles, abstracts, and keywords containing phrases "activity" AND ("recognition" OR "estimation" OR "classification") AND ("smartphone" OR "cell phone" OR "mobile phone"). The search was limited to full-length journal articles written in English. After removing duplicates, we read the titles and abstracts of the remaining publications. Studies that did not investigate HAR approaches were excluded from further screening. We then filtered out studies that employed auxiliary equipment, like wearable or ambient devices, and studies that required carrying multiple smartphones. Only studies that made use of commercially available consumer-grade smartphones (either personal or loaner) were read in full. We excluded studies that used the smartphone microphone or video camera for activity classification as they might record information about an individual's surroundings, including information about unconsented individuals, and thus hinder large-scale application of the approach due to privacy concerns. To focus on studies that mimicked free-living settings, we excluded studies that utilized devices strapped or glued to the body in a fixed position.

## 3. Results

Our search resulted in 1901 hits for the specified search criteria (Figure 2). After removal of articles that did not discuss HAR algorithms (n=793), employed additional hardware (n=150), or utilized microphones, cameras, or body-affixed smartphones (n=149), there were 108 references included in this review.

(**Figure 2** About here)

Most HAR approaches consist of four stages: *data acquisition*, *data preprocessing*, *feature extraction*, and *activity classification* (Figure 3). Here, we provide an overview of these steps and briefly point to significant methodological differences among the reviewed studies for each step. Table 1 summarizes specific aspects of each study. Of note, we decomposed data acquisition processes into sensor type, experimental environment, investigated activities, and smartphone location; we indicated which studies preprocessed collected measurements using signal correction methods, noise filtering techniques, and sensor orientation-invariant transformations; we marked investigations based on the types of signal features they extracted, as well as the feature selection approaches used; we indicated the adopted activity classification principles, utilized classifiers, and practices for accuracy reporting; and finally, we highlighted efforts supporting reproducibility and generalizability of the research. Before diving into these technical considerations, we first provide a brief description of study populations.

(**Figure 3** About here)

(**Table 1** About here)

## 3.1. Study populations

We use the term *study population* to refer to the group of individuals investigated in any given study. In the reviewed studies, data were usually collected from fewer than 30 individuals, although one larger study analyzed data from 440 healthy individuals [17]. Studies often included healthy adults in their twenties and thirties, with only a handful studies involving older individuals. Most studies did not report the full distribution of ages, only the mean age or the age range of participants (Figure 4). To get a sense of the distribution of participant ages, we attempted to reconstruct an overall approximate age distribution by assuming that the participants in each study are evenly distributed in age between the minimum and maximum ages, which may not be the case. A comparison of the reconstructed age distribution of study participants with nationwide age distributions clearly demonstrates that future HAR research in health settings needs to broaden the age spectrum of the participants. Less effort was devoted in the studies to investigating populations with different demographic and disease characteristics, such as elders [18–20] and individuals with Parkinson's disease [21].

(**Figure 4** About here)

## 3.2. Data acquisition

We use the term *data acquisition* to refer to a process of collecting and storing raw sub-second level smartphone measurements for the purpose of HAR. The data are typically collected from individuals by an application that runs on the device and samples data from built-in smartphone sensors according to a predefined schedule. We carefully examined the selected literature for details on the investigated population, measurement environment, performed activities, and smartphone settings.

In the reviewed studies, data acquisition typically took place in a research facility and/or nearby outdoor surroundings. In such environments, study participants were asked to perform a series of activities along predefined routes and to interact with predefined objects. The duration and order of performed activities were usually determined by the study protocol and the participant was supervised by a research team member. A less common approach involved observation conducted in free-living environments, where individuals performed activities without specific instructions. Such studies were likely to provide more insight into diverse activity patterns due to individual habits and unpredictable real-life conditions. Compared to a single laboratory visit, studies conducted in free-living environments also allowed investigators to monitor behavioral patterns over many weeks [22] or months [23].

Activity selection is one of the key aspects of HAR. The studies in our review tended to focus on a small set of activities, including sitting, standing, walking, running, and stair climbing. Less common activities involved various types of mobility, locomotion, fitness, and household routines. For instance, Wu et al. [24] differentiated between slow, normal, and brisk walking; Guvensan et al. [25] investigated multiple transportation modes, such as by car, bus, tram, train, metro, and ferry; Pei et al. [26] recognized sharp body-turns; and Della Mea et al. [27] looked into household activities, like sweeping a floor or walking with a shopping bag. More recent studies concentrated solely on walking recognition [28,29]. As shown in Table 1, the various measured activities in the reviewed studies can be grouped into classes: “posture” refers to lying, sitting, standing, or any pair of these activities; “mobility” refers to walking, stair climbing, body turns, riding an elevator or escalator, running, cycling, or any pair of these activities; “locomotion” refers to motorized activities; and “other” refers to various household and fitness activities or singular actions beyond the described groups.

(**Figure 5** About here)

The spectrum of investigated activities determines the choice of sensors used for data acquisition. At the time of writing, a standard smartphone is equipped with a number of built-in hardware sensors and protocols that can be used for activity monitoring, including an accelerometer, gyroscope, magnetometer, GPS, proximity sensor, and light sensor, as well as to collect information on ambient pressure, humidity, and temperature (Figure 5). Our literature review revealed that the most commonly used sensors for HAR are the accelerometer, gyroscope, and magnetometer, which capture data about acceleration, angular velocity, and phone orientation, respectively, and provide temporally dense, high-resolution measurements for distinguishing among activity classes. Inertial sensors were often used synchronously to provide more insight into the dynamic state of the device. Some studies showed that the use of a single sensor can yield similar accuracy of activity recognition as using multiple sensors in combination [30]. To alleviate the impact of sensor position, some researchers collected data using the built-in barometer and GPS sensors to monitor changes in altitude and geographic location [31–33]. Certain studies benefited from using the broader set of capabilities of smartphones; for example, some researchers additionally exploited the proximity sensor and light sensor to allow recognition of a measurement's context, e.g., the distance between a smartphone and the individual's body, and changes between in-pocket and out-of-pocket locations based on changes in illumination [34,35]. The selection of sensors was also affected by secondary research goals, such as simplicity of classification and minimization of battery drain. In these studies, data acquisition was carried out using a single sensor (e.g., accelerometer [22]), a small group of sensors (e.g., accelerometer and GPS [36]), or a purposely modified sampling frequency or sampling scheme (e.g., alternating between data collection and non-collection cycles) to reduce the volume of data collected and processed [37].

(**Figure 6** About here)

Sampling frequency specifies how many observations are collected by a sensor within a one-second time interval. The selection of sampling frequency is usually performed as a trade-off between measurement accuracy and battery drain. Sampling frequency in the reviewed studies typically ranged between 20 and 30Hz for inertial sensors and 1 and 10Hz for the barometer and GPS. The most significant variations were seen in studies where limited energy consumption was a priority (e.g., accelerometer sampled at 1Hz [38]) or if investigators used advanced signal processing methods, such as time-frequency decomposition methods, or activity templates that required higher frequency sampling (e.g., accelerometer sampled at 100Hz [39]).

A crucial parameter in the data acquisition process is the smartphone's location on the body. This is important mainly because of the nonstationary nature of real-life conditions and the strong effect it has on the smartphone's inertial sensors. The main challenge in HAR in free-living conditions is that data recorded by the accelerometer, gyroscope, and magnetometer sensors differ between the upper and lower body as the device is not affixed to any specific location or orientation [40]. Therefore, it is essential that studies collect data from as many body locations as possible to ensure generalizability of results. In the reviewed literature, study participants were often instructed to carry the device in a pants pocket (either front or back), although a number of studies also considered other placements, such as jacket pocket [41], bag or backpack [42,43], and holding the smartphone in the hand [44] or in a cupholder [45].

To establish the ground truth for physical activity in HAR studies, data were usually annotated manually by trained research personnel or by the study participants themselves [46,47]. However, we also noted several approaches that automated this process both in controlled and free-living conditions. For instance, Liang et al. [22] labelled data using a designated smartphone application, while Cruciani et al. [48], used a built-in step counter and GPS data to produce "weak" labels. The annotation was also done using the built-in microphone [49], video camera [18,20], or an additional body-worn sensor [29].

Finally, the data acquisition process in the reviewed studies was carried out on purposely designed applications which captured data. In studies with *online* activity classification, the collected data did not leave the device but instead the entire HAR pipeline was implemented on the smartphone; in contrast, studies using *offline* classification transmitted data to an external (remote) server for processing using a cellular, Wi-Fi, Bluetooth, or wired connection.

### 3.3. Data preprocessing

We use the term *data preprocessing* to refer to a collection of procedures aimed at repairing, cleaning, and transforming measurements recorded for HAR. The need for such step is threefold: (1) measurement systems embedded in smartphones are often less stable than research-grade data acquisition units, and the data might therefore be sampled unevenly or there might be missingness or sudden spikes that are unrelated to an individual's actual behavior; (2) the spatial orientation (how the phone is situated in a person's pocket, say) of the device influences tri-axial measurements of inertial sensors, thus potentially degrading the performance of the HAR system; and (3) despite careful planning and execution of the data acquisition stage, data quality may be compromised due to other unpredictable factors, e.g., lack of compliance by the study participants, unequal duration of activities in the measurement (i.e., dataset imbalance), or technological issues.

In our literature review, the first group of obstacles was typically addressed using signal processing techniques (in Table 1, see "standardization"). For instance, to alleviate the mismatch between requested and effective sampling frequency, Derawi and Bours [50] proposed the use of linear interpolation, while Gu et al. [51] utilized spline interpolation (Figure 7). Such procedures were imposed on a range of affected sensors, typically the accelerometer, gyroscope, magnetometer, and barometer. Further time-domain preprocessing considered data trimming, carried out to remove unwanted data components. For this purpose, the beginning and end of each activity bout, a short period of activity of a specified kind, were clipped as nonrepresentative for the given activity [41]. During this stage, the researchers also dealt with dataset imbalance, which occurs when there are different numbers of observations for different activity classes in the training dataset. Such a situation makes the classifier susceptible to overfitting in favor of the larger class; in the reviewed studies, this issue was resolved using up-sampling or down-sampling of data [17,52–54]. Additionally, the measurements were processed for high-frequency noise cancellation (i.e., "denoising"). The literature review identified several methods suitable for this task, including the use of low-pass finite impulse response filters (with cut-off frequency typically equal to 10Hz for inertial sensors and 0.1Hz for barometers) [55,56], which remove the portion of the signal that is unlikely to result from the activities of interest; weighted moving average [50]; moving median [40,57]; and singular value decomposition [58]. GPS data were sometimes de-noised based on the maximum allowed positional accuracy [59].

Another element of data preprocessing considers device orientation (in Table 1, see “transformation”). Smartphone measurements are sensitive to device orientation, which may be due to clothing, body shape, and movement during dynamic activities [52]. One of the popular solutions reported in the literature was to transform the three-dimensional signal into a univariate vector magnitude that is invariant to rotations and more robust to translations. This procedure was often applied to accelerometer, gyroscope, and magnetometer data. Accelerometer data was also subjected to digital filtering by separating the signal into linear (related to body motions) and gravitational (related to device spatial orientation) acceleration [60]. This separation was typically performed using high-pass Butterworth filter of low order (e.g., order 3) with a cut-off frequency below 1Hz. Other approaches transformed tri-axial into bi-axial measurement with horizontal and vertical axes [44], or projected the data from the device coordinate system into a fixed coordinate system (e.g., coordinate system of a smartphone that lies flat on the ground) using a rotation matrix (Euler angle-based [61] or quaternion [42,62]).

(**Figure 7** About here)

## 3.4. Feature extraction

We use the term *feature extraction* to refer to a process of selecting and computing meaningful summaries of smartphone data for the goal of activity classification. A typical extraction scheme includes data visualization, data segmentation, feature selection, and feature calculation. A careful feature extraction step allows investigators not only to understand the physical nature of activities and their manifestation in digital measurements, but also, and more importantly, to help uncover hidden structures and patterns in the data. The identified differences are later quantified through various statistical measures to distinguish between activities. In an alternative approach, the process of feature extraction is automated using deep learning, which handles feature selection using simple signal processing units, called neurons, that have been arranged in a network structure that is multiple layers deep [54,63–65]. As with many applications of deep learning, the results may not be easily interpretable.

The conventional approach to feature extraction begins with data exploration. For this purpose, researchers in our reviewed studies employed various graphical data exploration techniques like scatter plots, lag plots, autocorrelation plots, histograms, and power spectra [66]. The choice of tools was often dictated by the study objectives and methods. For example, research on inertial sensors typically presented raw three-dimensional data from accelerometers, gyroscopes, and magnetometers plotted for the corresponding activities of standing, walking, and stair climbing [45,67,68]. Acceleration data were often inspected in the frequency domain, particularly to observe periodic motions of walking, running, and cycling [40], and the impact of external environment, like natural vibration frequencies of a bus or a subway [69]. Locomotion and mobility were investigated using estimates of speed derived from GPS. In such settings, investigators calculated the average speed of the device and associated it with either the group of motorized (car, bus, train, etc.) or non-motorized (walking, cycling, etc.) modes of transportation.

In the next step, measurements are divided into smaller fragments (also, segments or epochs) and signal features are calculated for each fragment (Figure 8). In our reviewed studies, this segmentation was typically conducted using a windowing technique that allows consecutive windows to overlap. The window size usually had a fixed length that varied from 1 to 5s, while the overlap of consecutive windows was often set to 50%. Several studies investigated how to select the optimal window size, which emphasizes the importance of this parameter to the performance of HAR systems [70–72]. This calibration aims to closely match the window size with the duration of a single instance of the activity (e.g., one step). Similar motivation led researchers to seek more adaptive segmentation methods. One idea was to segment data based on specific time-domain events, like zero-cross points (when signal changes value from positive to negative or vice versa), peak points (local maxima), or valley points (local minima), which represent the start and end points of a particular activity bout [50,52]. This allowed for segments to have different lengths corresponding to a single fundamental period of the activity in question. Such an approach was typically used to recognize quasiperiodic activities like walking, running, and stair climbing [58].

(**Figure 8** About here)

The literature described a large variety of signal features used for HAR, which can be divided into several categories based on the initial signal processing procedure. This enables one to distinguish between activity templates (i.e., raw signal), deep features (i.e., hidden features calculated within layers of deep neural networks), time-domain features (i.e., statistical measures of time series data), and frequency-domain features (i.e., statistical measures of frequency representation of time series data). The most popular features in the reviewed papers were calculated from time-domain signals as descriptive statistics, such as local mean, variance, minimum and maximum, interquartile range, signal energy (defined as the area under the squared magnitude of the considered continuous signal), and higher order statistics. Other time-domain features included mean absolute deviation, mean (or zero) crossing rate, regression coefficients, and autocorrelation. Some studies described novel and customized time-domain features, like histograms of gradients [73], and the number of local maxima and minima, their amplitude, and the temporal distance between them [37]. Time-domain features were typically calculated over each axis of the three-dimensional measurement or orientation-invariant vector magnitude. Studies that used GPS also calculated average speed [59,74,75], while studies that used the barometer analyzed the pressure derivative [76].

Signals transformed to the frequency domain were less exploited in the literature. A commonly performed signal decomposition used the fast Fourier transform (FFT) [77,78], an algorithm that converts a temporal sequence of samples to a sequence of frequencies present in that sample. The essential advantage of frequency-domain features over time-domain features is their ability to identify and isolate certain periodic components of performed activities. This enabled researchers to estimate (kinetic) energy within particular frequency bands associated with human activities, like gait and running [46], as well as with different modes of locomotion [69]. Other frequency-domain features included spectral entropy and parameters of the dominant peak, e.g., its frequency and amplitude.

Activity templates function essentially as blueprints for different types of physical activity. In the HAR systems we reviewed, these templates were compared to patterns of observed raw measurements using various distance metrics [36,79], such as the Euclidean or Manhattan distance. Given the heterogeneous nature of human activities, activity templates were often enhanced using techniques similar to dynamic time warping [29,52], which measures the similarity of two temporal sequences that may vary in speed. As an alternative to raw measurements, some studies used signal symbolic approximation, which translates a segmented time series signal into sequences of symbols based on a predefined mapping rule (e.g., amplitude between -1 and -0.5g represents symbol “a”, amplitude between -0.5 and 0g represents symbol “b”, and so on) [80–82].

In the reviewed articles, the number of extracted features typically varied from a few to a dozen. However, some studies purposely calculated too many features (sometimes hundreds) and let the analytical method perform variable selection, i.e., identify those features that were most informative for HAR. Support vector machines [76,83], gain ratio [84], recursive feature elimination [36], correlation-based feature selection [46], and principal component analysis [85] were among the popular feature selection/dimension reduction methods used. A comparison of feature selection methods is provided by Saeedi and El-Sheimy [86].

## 3.5. Activity classification

We use the term *activity classification* to refer to a process of associating extracted features with particular activity classes based on the adopted classification principle. The classification is typically performed by a supervised learning algorithm that has been trained to recognize patterns between features and labeled physical activities in the training dataset. The fitted model is then validated on separate observations, using a validation dataset, usually data obtained from the same group of study participants. The comparison between predictions made by the model and the known true labels allows one to assess the accuracy of the approach. This section summarizes the methods used in classification and validation, and also provides some insights into reporting on HAR performance.

The choice of classifier aims to identify a method that has the highest classification accuracy for the collected datasets and for the given data processing environment (e.g., online vs. offline). The reviewed literature included a broad range of classifiers, from simple decision trees [18], k-nearest neighbors [60], support vector machines [87–89], logistic regression [21], naïve Bayes [90], and fuzzy logic [59] to ensemble classifiers such as random forest [71], XGBoost [91], AdaBoost [40,92], bagging [24], and deep neural networks [43,55,77,93–95]. Simple classifiers were frequently compared to find the best solution in the given measurement scenario [48,84,96–98]. A similar type of analysis was implemented for ensemble classifiers [74]. Incremental learning techniques were proposed to adapt the classification model to new data streams and unseen activities [99,100]. To increase the effectiveness of HAR, some studies used a hierarchical approach, where the classification was performed in separate stages and each stage could use a different classifier. The multi-stage technique was used for gradual decomposition of activities (coarse-grained first, then fine-grained) [22,35,47,55] and to handle the predicament of changing sensor location (body location first, then activity) [87]. Classification accuracy could also be improved by using post-processing, which relies on modifying the initially assigned label using the rules of logic and probability. The correction was typically performed based on activity duration [69], activity sequence [25], and activity transition probability and classification confidence [75,101].

The selected method is typically cross-validated, which splits the collected dataset into two or more parts—training and testing—and only uses the part of the data for testing that was not used for training. The literature mentions a few cross-validation procedures, with *k*-fold and leave-one-out cross-validation being the most common [102]. Popular train-test proportions were 90-10, 70-30, and 60-40. A validation is especially valuable if it is performed using studies with different demographics and smartphone use habits. Such an approach allows one to understand the generalizability of the HAR system to real-life conditions and populations. We found a few studies that followed this validation approach [18,21,66].

Activity classification is the last stage of HAR. In our review, we found that analysis results were typically reported in terms of the classification accuracy using various standard metrics like precision, recall, and F-score. Overall, the investigated studies reported very high classification accuracies, typically above 95%. Several comparisons revealed that ensemble classifiers tended to outperform individual or single classifiers [27,72], and deep learning classifiers tended to outperform both individual and ensemble classifiers [43]. More nuanced summaries used the confusion matrix, which allows one to examine which activities are more likely to be classified incorrectly. This approach was particularly useful for visualizing classification differences between similar activities, such as normal and fast walking or bus and train riding. Additional statistics were usually provided in the context of HAR systems designed to operate on the device. In this case, activity classification needed to be balanced among acceptable classifier performance, processing time, and battery drain [103]. The desired performance optimum was obtained by making use of dataset remodeling (e.g., by replacing the oldest observations with the newest ones), low-cost classification algorithms, limited preprocessing, and conscientious feature selection [40,81]. Computation time was sometimes reported for complex methods, such as deep neural networks [20,77,104] and extreme learning machine [105], as well as for symbolic representation [80,81] and in comparative analyses [41]. A comprehensive comparison of results was difficult or impossible, as discussed below.

## 4. Discussion

Over the past decade, many studies have investigated HAR using smartphones. The reviewed literature provides detailed descriptions of essential aspects of data acquisition, data preprocessing, feature extraction, and activity classification. Studies were conducted with one or more objectives, e.g., to limit technological imperfections (e.g., no GPS signal reception indoors), to minimize computational requirements (e.g., for online processing of data directly on the device), and to maximize classification accuracy (all studies). Our review summarizes the most frequently used methods and offers available alternatives.

As expected, no single activity recognition procedure was found to work in all settings, which underlines the importance of designing methods and algorithms that address specific research questions in health while keeping the specifics of study cohort in mind (e.g., age distribution, extent of device use, and nature of disability). While datasets were usually collected in laboratory settings, there was little evidence that algorithms trained using data collected in these controlled settings could be generalized to free-living conditions [106,107]. In free-living settings, duration, frequency, and specific ways of performing any activity are subject to context and individual ability, and these degrees of freedom need to be considered in the development of HAR systems. Validation of these data in free-living settings is essential, as the true value of HAR systems for public health will come through transportable and scalable applications in large, long-term observational studies or real-world interventions.

Some studies were conducted with a small number of able-bodied volunteers. This makes the process of data handling and classification easier but also limits the generalizability of the approach to more diverse populations. The latter point was well demonstrated by del Rosario et al. [18] and Albert et al. [21]. In the first study, the authors observed that the performance of a classifier trained on a young cohort significantly decreases if validated on an older cohort. Similar conclusions can be drawn from the second study, where the observations on healthy individuals did not replicate in individuals with Parkinson's disease.

The majority of the studies we reviewed utilized stationary smartphones at a single body position (i.e., a specific pant pocket), sometimes even with a fixed orientation. However, such scenarios are rarely observed in real-life settings, and these types of studies should be considered more as proofs of concept. Many studies provided only incomplete descriptions of the experimental setup and study protocol, and provided few details on demographics, environmental context, and the details of the performed activities. Such information should be reported as fully and accurately as possible.

Only a few studies considered classification in a context that involves activities outside the set of activities the system was trained on; for example, if the system was trained to recognize walking and running, these were the only two activities that the system was later tested on. However, real-life activities are not limited to a prescribed set of behaviors, i.e., we do not just sit still, stand still, walk, and climb stairs. These classifiers, when applied to free-living conditions, will naturally miss the activities they were not trained on but will also likely overestimate those activities they were trained on. An improved scheme could assume that the observed activities are a sample from a broader spectrum of possible behaviors or assess the uncertainty associated with the classification of each type of activity [79].This could also provide for an adaptive approach that would enable observation / interventions suited to a broad range of activities relevant for health, including decreasing sedentary behavior, increasing active transport (i.e., walking, bicycling, or public transit), and improving circadian patterns / sleep.

Use of personal digital devices, in particular smartphones, makes it possible to follow large numbers of individuals over long periods of time, but invariably investigators need to consider approaches to missing sensor data, which is a common problem. The importance of this problem is illustrated in a recent paper that introduced a resampling approach to imputing missing smartphone GPS data; the authors found that relative to linear interpolation—the naïve approach to missing spatial data—imputation resulted in a 10-fold reduction in the error averaged across all daily mobility features [108]. On the flip side of missing data is the need to propagate uncertainty, in a statistically principled way, from the gaps in the raw data to the inferences that investigators wish to draw from the data. It is a common observation that different people use their phones differently, and some may barely use their phones at all; the net result is not that the data collected from these individuals are not useful, but instead the data are less informative about the behavior of this individual than they ideally might be. Dealing with missing data and accounting for the resulting uncertainty is important because it means that one does not have to exclude participants from a study because their data fail meet some arbitrary threshold of completeness; instead, everyone counts and every bit of data from each individual counts.

(**Table 2** About here)

Collection of behavioral data using smartphones understandably raises concerns about privacy; however, investigators in health research are well-positioned to understand and address these concerns given that health data are generally considered personal and private in nature. Consequently, there are established practices and common regulations on human subjects' research, where informed consent of the individual to participate is one of the key foundations of any ethically conducted study. Federated learning is a machine learning technique that can be used to train an algorithm across decentralized devices, here smartphones, using only local data (data from the individual) and without the need to exchange data with other devices. This approach appears at first to provide a powerful solution to the privacy problem: the personal data never leave the person's phone and only the outputs of the learning process, generally parameter estimates, are shared with others. This is where the tension between privacy and the need for reproducible research arises, however. The reason for data collection is to produce generalizable knowledge, but according to an often-cited study, 65 percent of medical studies were inconsistent when retested and only 6 percent were completely reproducible [12]. In the studies reviewed here, only 5 out of 108 made the source code or the methods used in the study publicly available. For a given scientific question, studies that are not replicable require the collection of more private and personal data; this highlights the importance of reproducibility of studies, especially in health, where there are both financial and ethical considerations when conducting research. If federated learning provides no possibility to confirm data analyses, to re-analyze data using different methods, or to pool data across studies, it by itself cannot be the solution to the privacy problem. Nevertheless, the technique may act as inspiration for developing privacy-preserving methods that also enable future replication of studies. One possibility is to use publicly available datasets (Table 2). If sharing of source code were more common, HAR methods could be tested on these publicly available datasets, perhaps in a similar way as datasets of handwritten digits are used to test classification methods in machine learning research. Although some efforts have been made in this area [113,114], the recommended course of action assumes collecting and analyzing data from a large spectrum of sensors on diverse and under-studied populations and validating classifiers against widely accepted gold standards.

When accurate, reproducible, and transportable methods coalesce to recognize a range of relevant activity patterns, smartphone-based HAR approaches will provide a fundamental tool for public health researchers and practitioners alike. We hope that this paper has provided to the reader some insights into how smartphones may be used to quantify human behavior in health research and the complexities that are involved in the collection and analysis of such data in this challenging but important field.

## Funding sources

Drs Straczkiewicz and Onnela are supported by NHLBI award U01HL145386 and NIMH award R37MH119194. Dr Onnela is also supported by NIMH award U01MH116928. Dr James is supported by NCI award R00CA201542 and NHLBI award R01HL150119.

## Author contribution

MS conducted the review, prepared figures, and wrote the initial draft. PJ and JPO revised the manuscript. JPO supervised the project. All authors reviewed the manuscript.

## Competing Interests

Authors declare no competing interests.

**Figures**

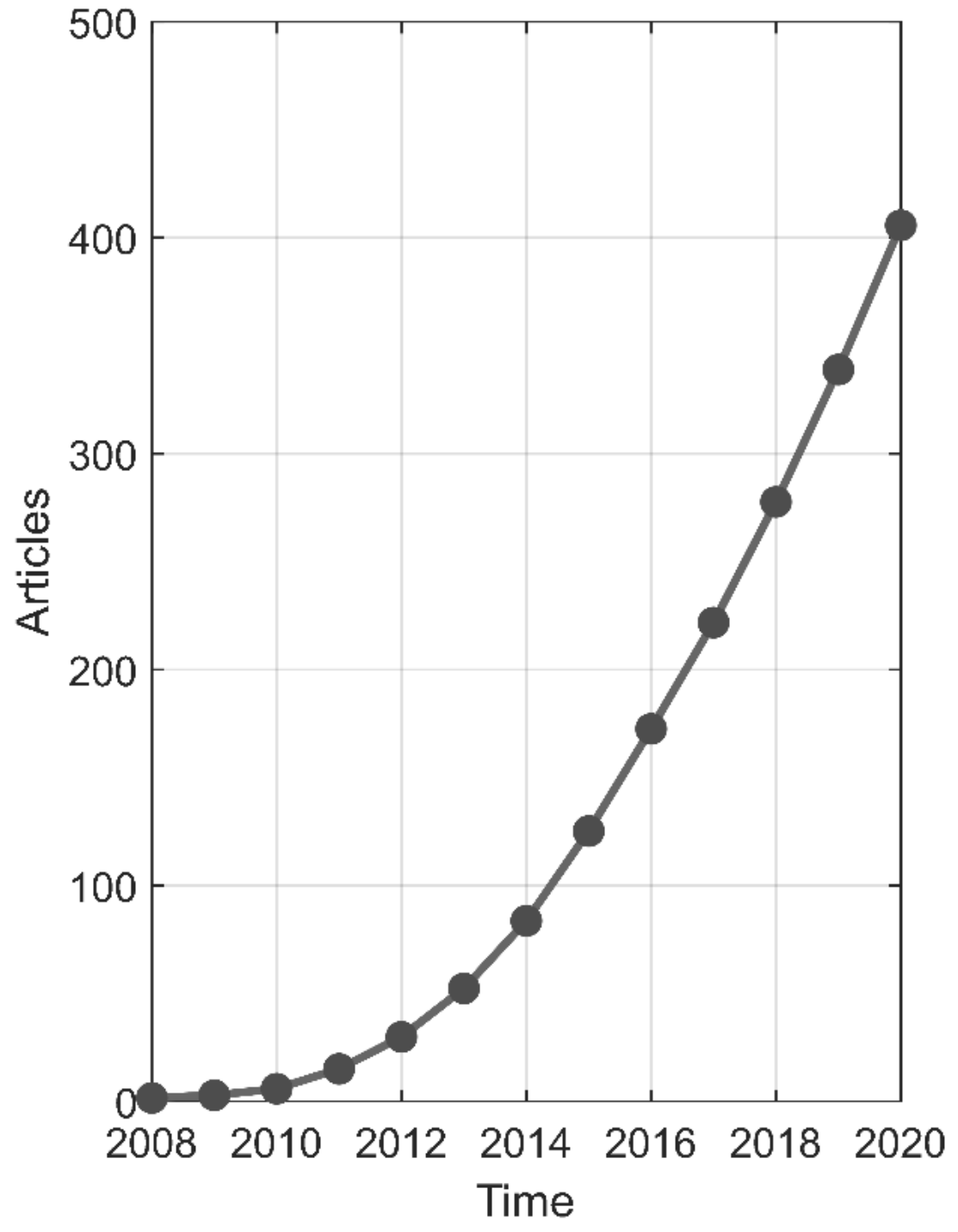

**Figure 1.** Cumulative number of peer-reviewed articles on human activity recognition (HAR) using smartphones published between January 2008 and December 2020, based on a search of PubMed, Scopus, and Web of Science databases (for details, see Methods).

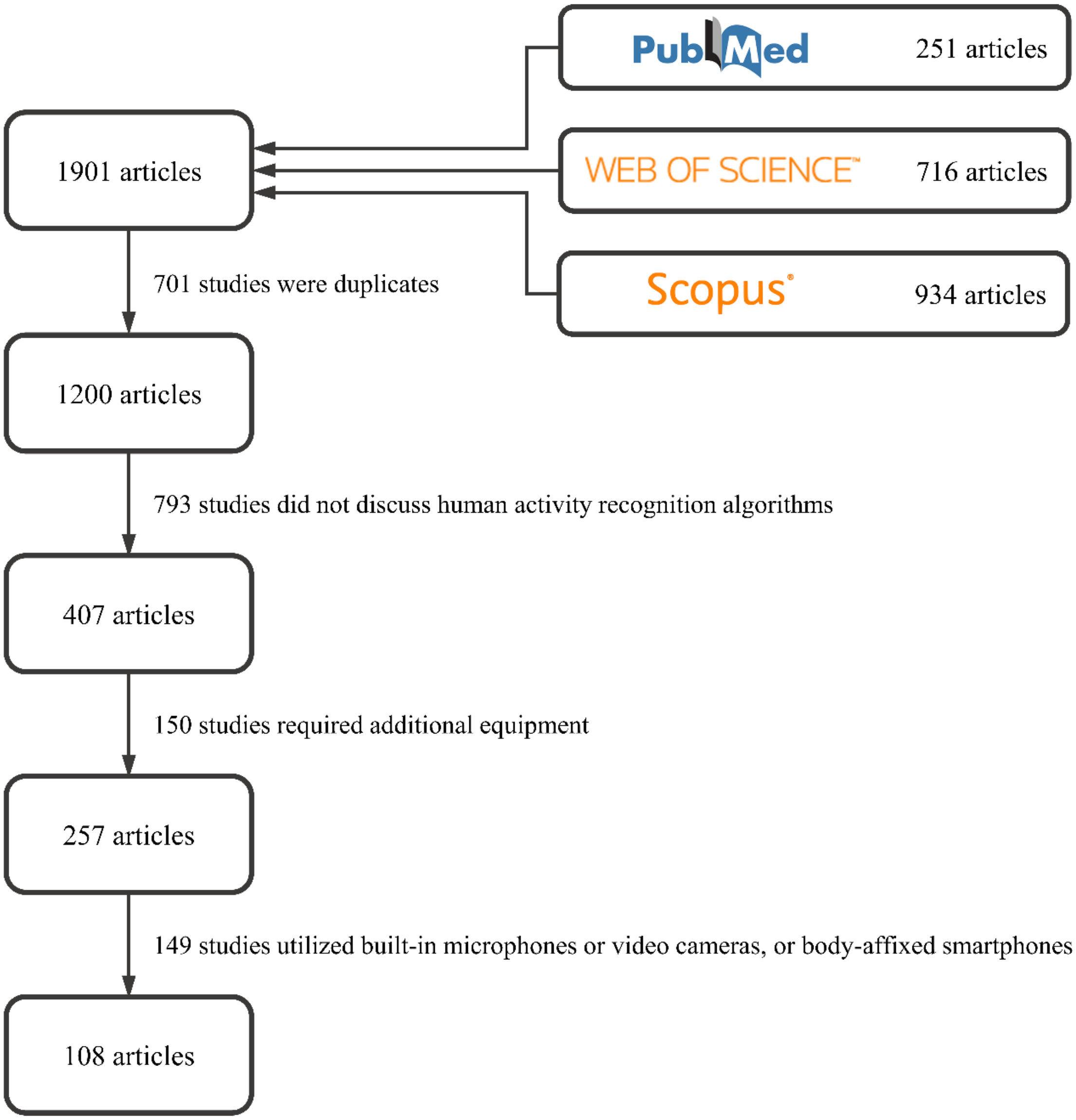

**Figure 2.** PRISMA diagram of the literature search process. The search was conducted in PubMed, Scopus, and Web of Science databases and included full-length peer-reviewed articles written in English. The search was carried out on January 2, 2021.

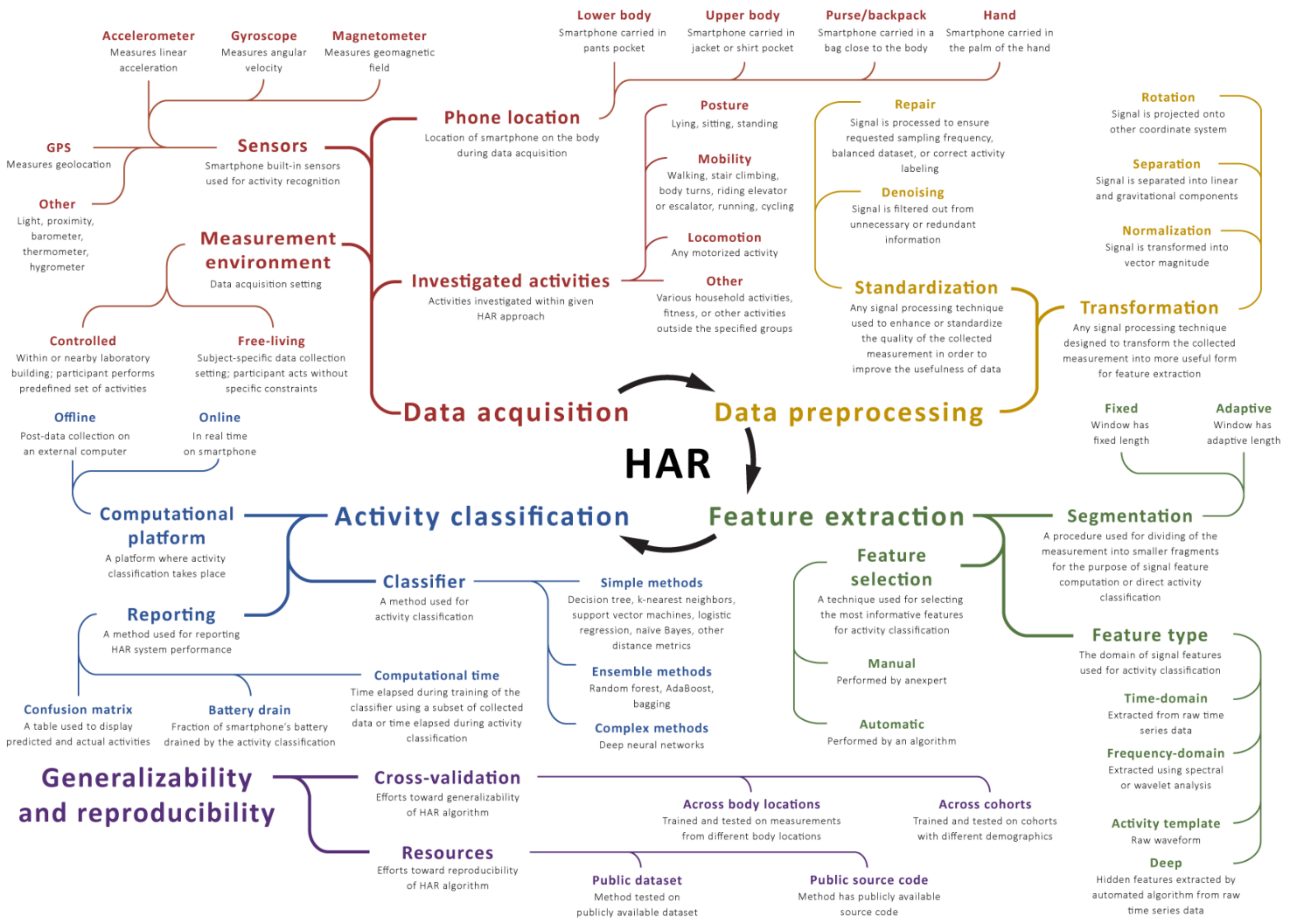

**Figure 3.** Map of concepts and operational definitions of terms used in smartphone-based human activity recognition (HAR) used in Table 1.

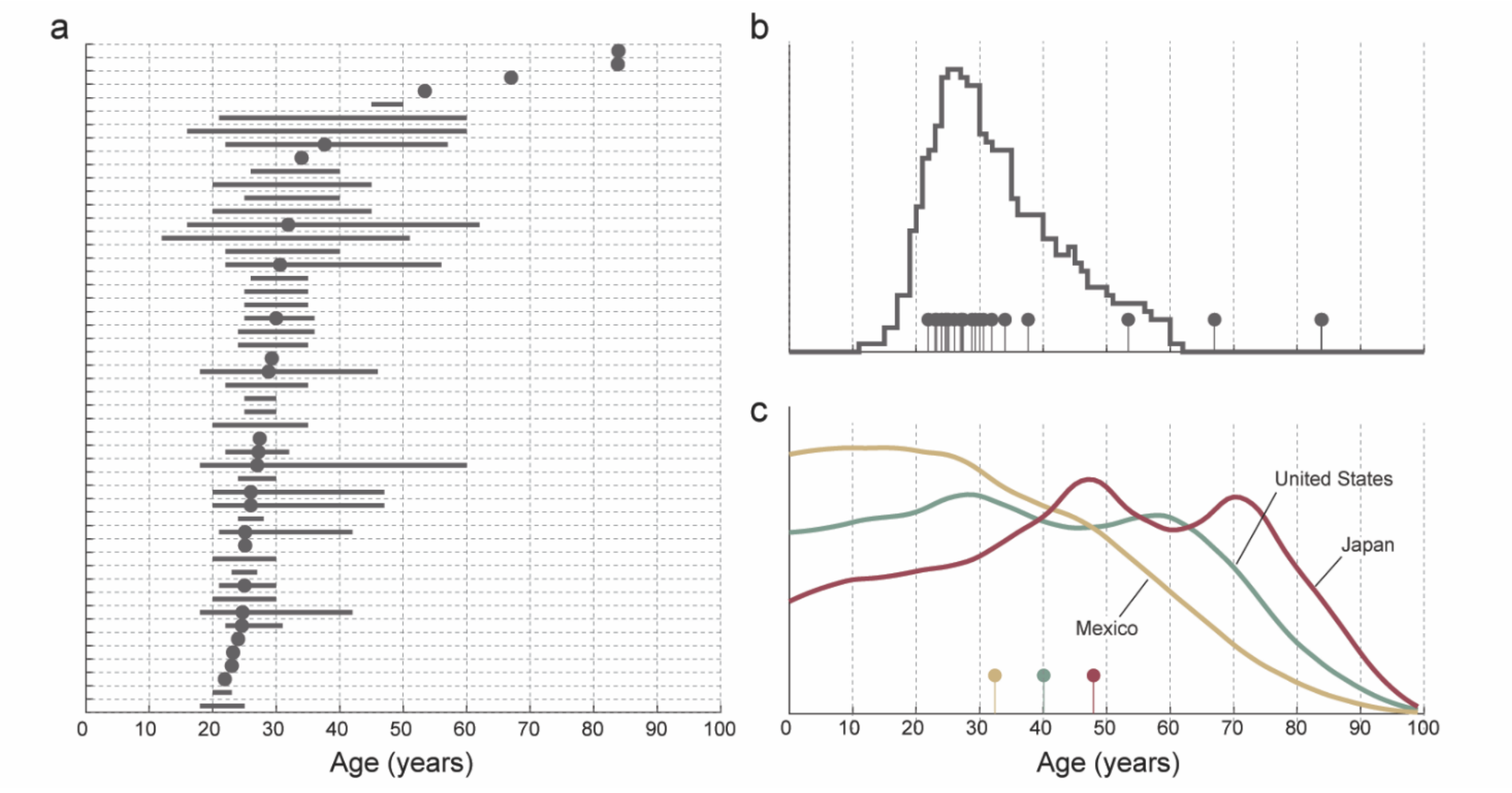

**Figure 4.** Age of populations examined in the reviewed papers (panel a) is typically described by its range (lines) or mean (dots). Each row corresponds to a study. Reconstructed age distribution (panel b) in the reviewed studies (see text). Nationwide age distributions (panel c) of three countries offer a stark contrast with the reconstructed distribution of study participant ages.

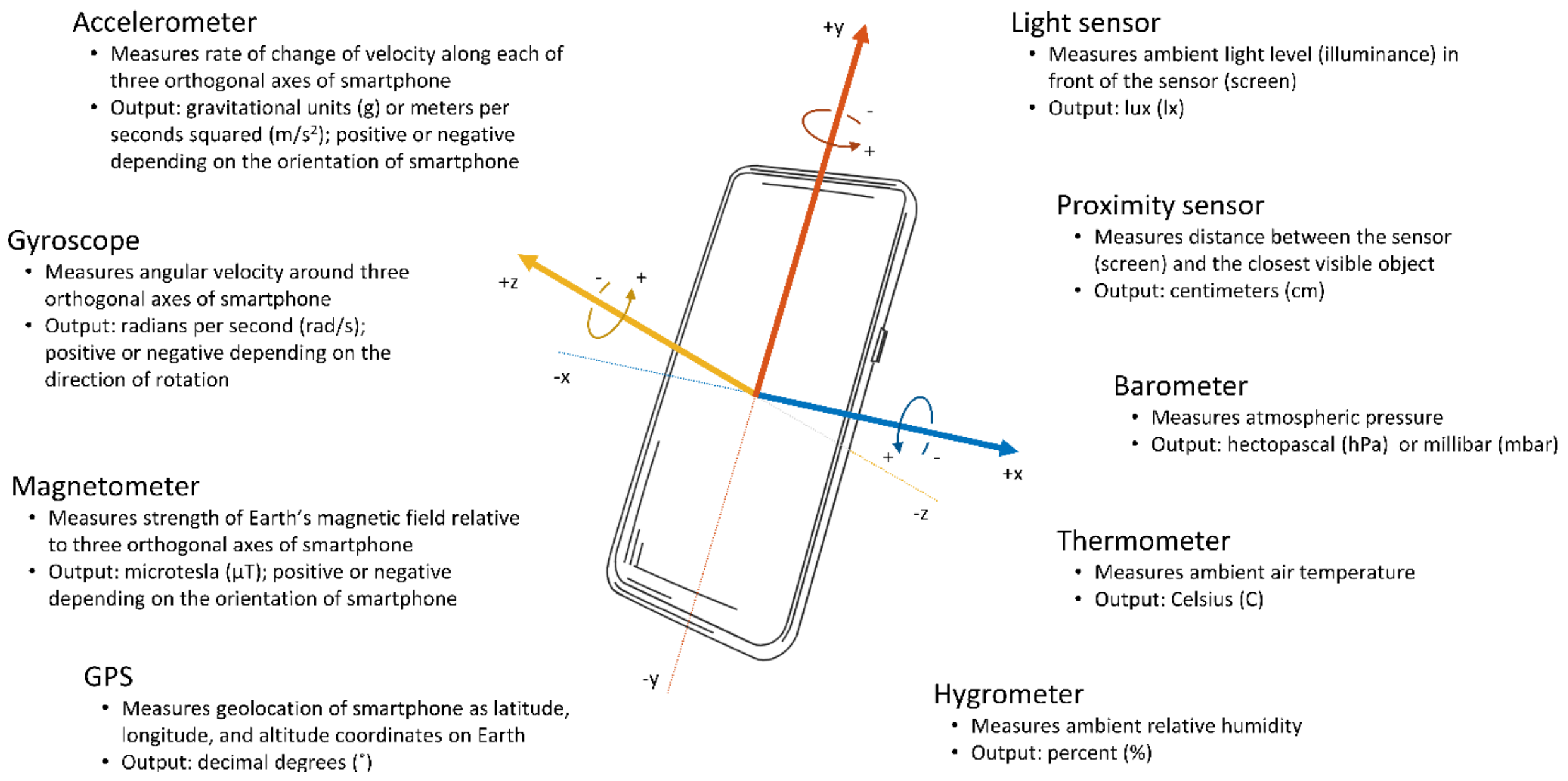

**Figure 5.** Standard smartphone sensors and their orientation with respect to the body of the phone.

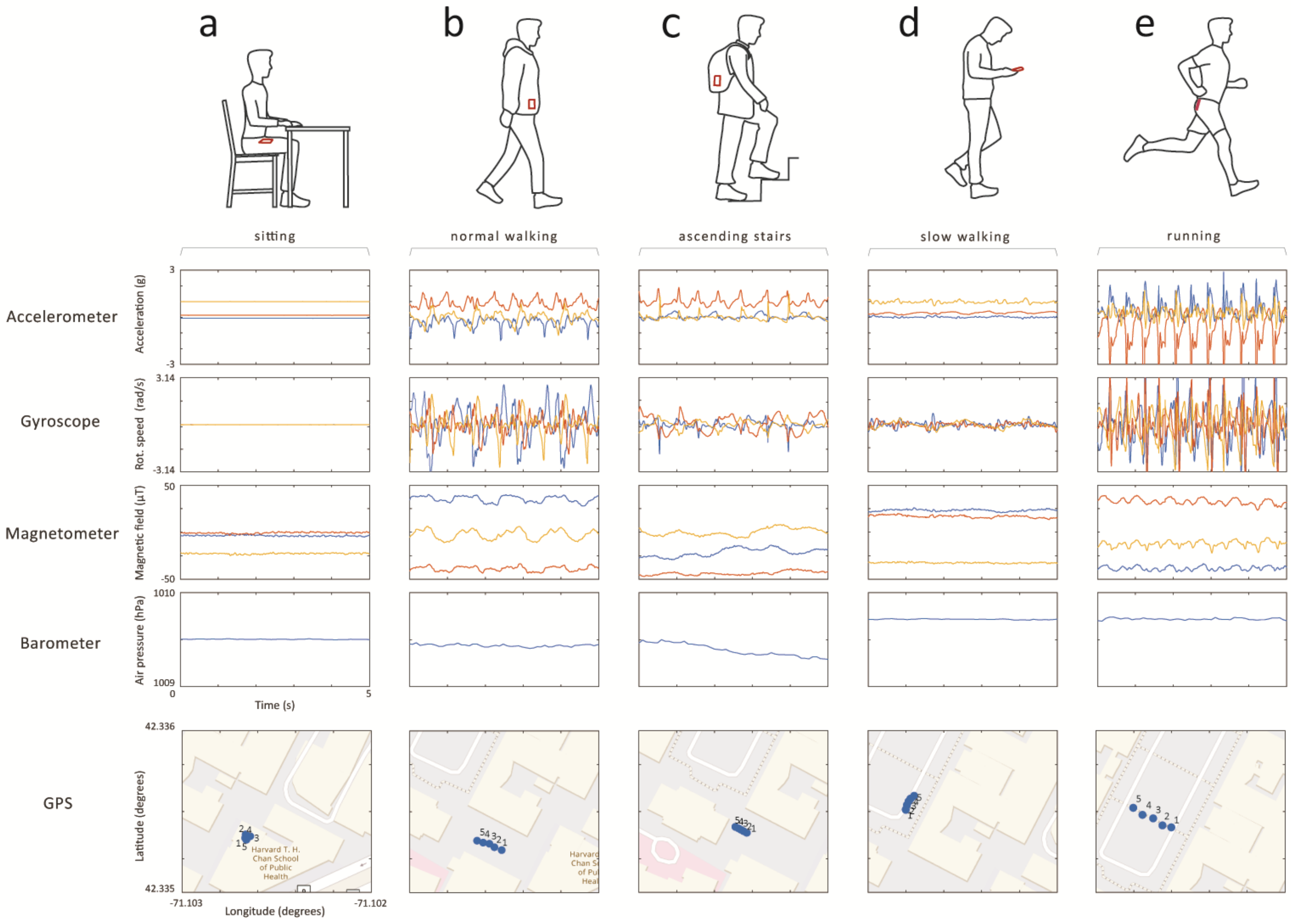

**Figure 6.** Examples of raw smartphone sensor data collected in a naturalistic setting. In panel a, a person is sitting by the desk with smartphone placed in the front pants pocket; in panel b, a person is walking normally (~1.9 steps per second) with smartphone placed in jacket pocket; in panel c, a person is ascending stairs with smartphone placed in the backpack; in panel d, a person is walking slowly (~1.4 steps per second) holding smartphone in hand; in panel e, a person is jogging (~2.8 steps per second) with smartphone placed in back shorts pocket.

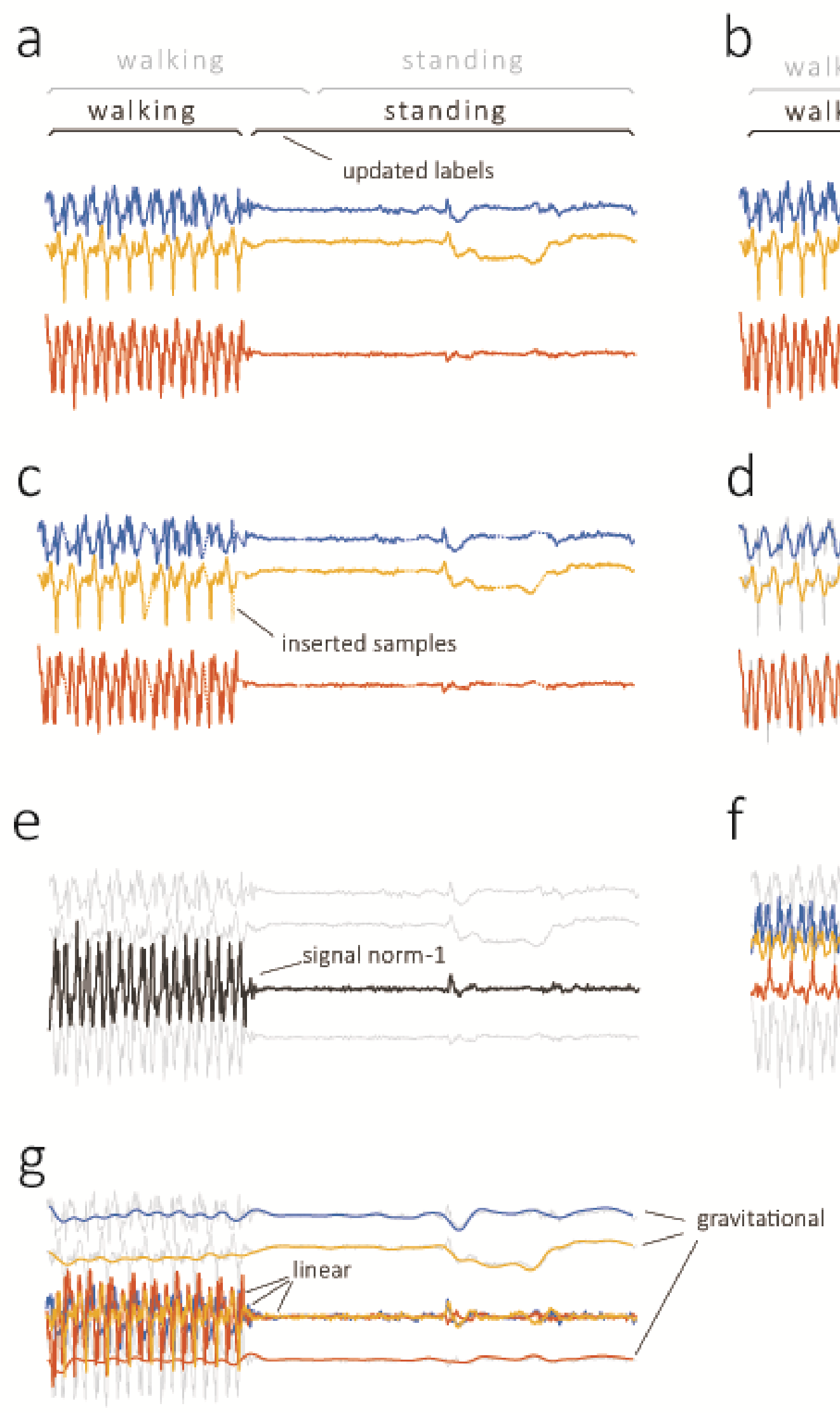

**Figure 7.** Common data preprocessing steps include standardization and transformation. Standardization includes relabeling (panel a), when labels are reassigned to better match transitions between activities; trimming (panel b), when part of the signal is removed to balance the dataset for system training; interpolation (panel c), when missing data are filled in based on adjacent observations; and de-noising (panel d), when the signal is filtered from redundant components. Transformation includes normalization (panel e), when the signal is normalized to unidimensional vector magnitude; rotation (panel f), when the signal is rotated to different coordinate system; and separation (panel g), when the signal is separated into linear and gravitational components. Raw accelerometer data are shown in grey and preprocessed data are shown using different colors.

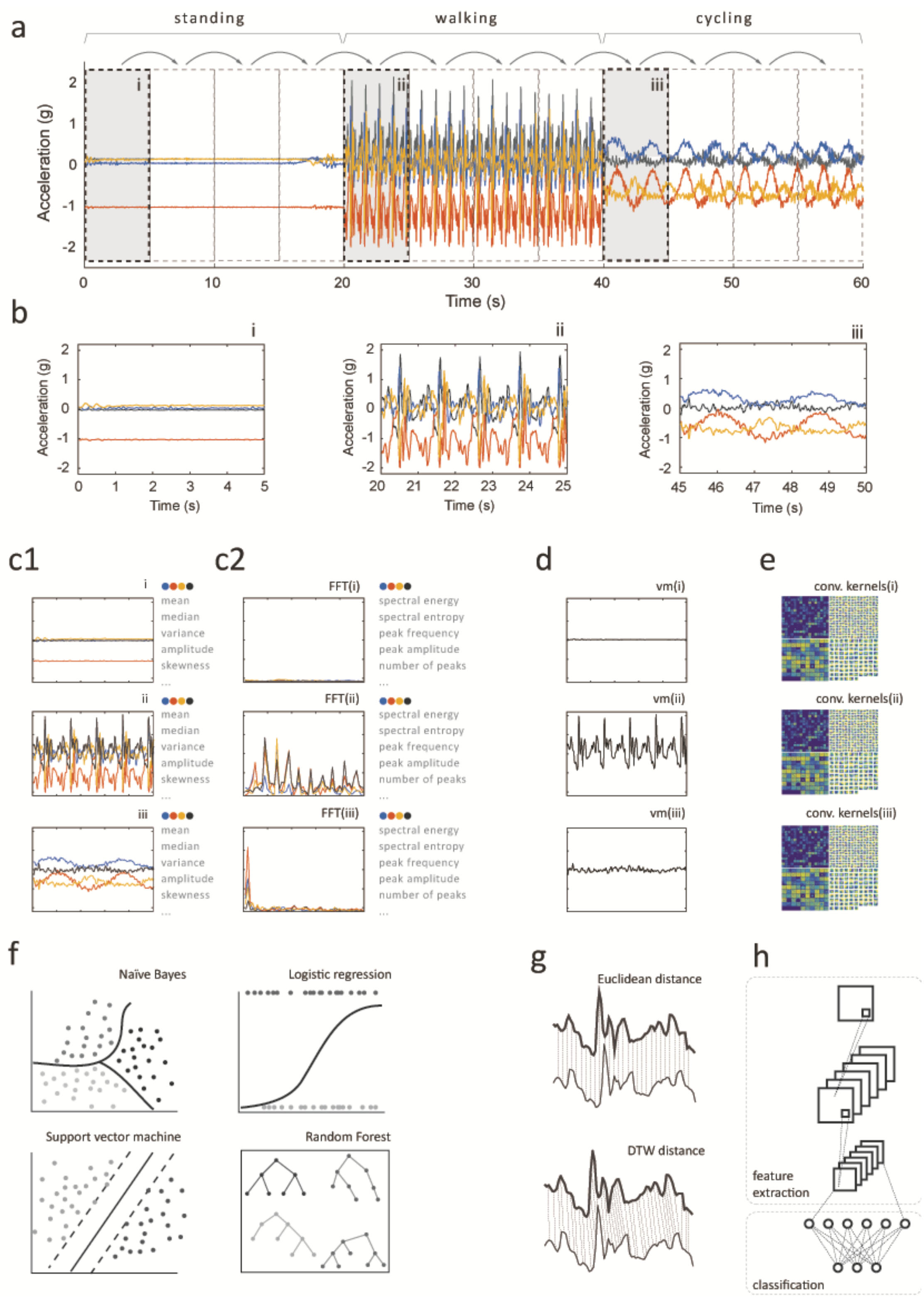

**Figure 8.** An analyzed measurement (panel a) is segmented into smaller fragments using a sliding window (panel b). Depending on the approach, each segment may then be used to compute time-domain (panel c1) or frequency-domain features (panel c2), but also it may serve as the activity template (panel d), or as input for deep learning networks that compute hidden ("deep") features (panel e). The selected feature extraction approach determines the activity classifier: time- and frequency-domain features are paired with machine learning classifiers (panel f) and activity templates are investigated using distance metrics (panel g), while deep features are computed within embedded layers of convolutional neural networks (panel h).

**Tables**

**Table 1.** Summary of HAR Systems Using Smartphones. The columns correspond to the 108 reviewed studies and the rows correspond to different technical aspects of each study. Cells marked with a cross (x) indicate that the given study used the given method, algorithm, or approach. Rows have been grouped to correspond to different stages of HAR, such as data processing, and color shading of rows indicates how frequently a particular aspect is present among the studies (darker shade corresponds to higher frequency).

| | | References | 46 | 125 | 24 | 21 | 50 | 66 | 126 | 26 | 47 | 18 | 99 | 60 | 87 | 103 | 30 | 56 | 90 | 41 | 22 | 76 |
|---|---|---|---|---|---|---|---|---|---|---|---|---|---|---|---|---|---|---|---|---|---|---|
| | | Publication year (20…) | 10 | 11 | 12 | 12 | 13 | 13 | 13 | 13 | 14 | 14 | 14 | 14 | 14 | 14 | 14 | 14 | 14 | 14 | 14 | 15 |
| Data acquisition | Measurement environment | Controlled | | x | x | x | x | x | x | x | x | x | x | x | x | | x | x | | x | | x |
| | | Free-living | x | | | | | | | | x | | | | | | | | | | x | |
| | Investigated activity | Posture | | x | x | x | | | | x | x | x | | x | x | x | x | x | | x | x | |
| | | Mobility | x | x | x | x | x | x | x | x | x | x | x | x | x | x | x | x | x | x | x | x |
| | | Locomotion | x | | | | | | | | x | | | | | | | | | | x | |
| | | Other | | x | | | | x | x | | | | | | | | | | | | x | |
| | Sensor | Accelerometer | x | x | x | x | x | x | x | x | x | x | x | x | x | x | x | x | x | x | x | x |
| | | Gyroscope | | | x | | | | | x | | x | | | | | x | x | x | | | x |
| | | Magnetometer | | | | | | | | x | | | | | | | x | x | x | | | x |
| | | GPS | x | | | | | | | | | | | | | | | x | x | | | |
| | | Other | | | | | | | | | | x | | | | | | x | | | | x |
| | Smartphone body location | Lower body | x | x | x | x | x | x | x | x | x | x | x | x | x | x | x | x | x | x | x | x |
| | | Upper body | | | | | | x | | | | | | | | | | | x | | | |
| | | Purse/backpack | x | | | | | | | | | | | | x | | | | x | | | |
| | | Hand | x | | | x | | | | | | | | | x | | | | x | x | | x |
| Data preprocessing | Standardization | Repair | x | | x | x | | | | | | | | | x | | | x | | x | x | |
| | | Denoising | | x | | | x | x | | | | x | | x | | | | x | | | | x |
| | Transformation | Normalization | x | x | x | | x | | | x | | | x | x | | x | x | | | | x | |
| | | Separation | | | | | | x | | x | | x | | x | | | | | | | | |
| | | Rotation | | | | | | | | x | | | | | | | | | | | | |
| Feature extraction | Segmentation | Fixed | x | x | x | x | | x | x | | | x | x | x | x | x | x | x | x | x | | x |
| | | Adaptive | | | | | x | | | | x | | | | | | | | | | x | |
| | Feature type | Time-domain | x | x | x | x | | x | x | x | x | x | x | x | x | x | x | x | x | x | x | x |
| | | Frequency-domain | x | | x | x | | x | x | x | x | | x | | x | | x | x | x | | x | |
| | | Deep | | | | | | | | | | | | | | | | | | | | |
| | | Activity template | | | | | x | | | | | | | | | | | | | | | |
| | Feature selection | Manual | | x | x | | x | x | x | x | x | x | x | | x | x | x | x | | x | x | |
| | | Automatic | x | | | x | | | | | | | | x | | | | x | x | | | x |
| Activity classification | Classifier | Simple | x | x | x | x | x | x | x | x | x | x | x | x | x | x | x | x | x | x | x | x |
| | | Ensemble | | | x | | | | | | | | | x | | | x | | | | | |
| | | Complex | | | | | | | | | | | | | | | | | | | | |
| | Computation platform | Offline | x | x | x | x | | | x | | x | | | x | x | | x | x | x | x | | x |
| | | Online | | | | | x | x | | x | | x | x | | | x | | | | | x | |
| | Reporting | Confusion matrix | | | x | x | x | x | x | x | | x | | x | x | | | | | | x | |
| | | Computational time | | | | | | | | | | | x | | | | | | x | x | x | |
| | | Battery drain | x | | | | | x | | x | | x | | | | x | | | | | x | |
| Generalizability and reproducibility | Cross-validation | Across cohorts | | | | x | | x | | | | x | | | x | | | | | | | |
| | | Across body locations | | | | | | | | | | | | | x | | | | | | | |
| | Resources | Public dataset | | | | | | | | | | | | x | | | | | | | | |
| | | Public source code | | | | | | | | | | | | | | | | | | | | |

| 38 | 111 | 86 | 37 | 34 | 127 | 100 | 32 | 44 | 128 | 35 | 129 | 74 | 61 | 31 | 58 | 130 | 96 | 70 | 109 | 59 | 42 | 84 | 49 | 36 | 85 | 69 | 40 | 27 | 52 | 104 | 80 | 39 | 48 | 72 | 51 | 131 | 71 | 132 | 101 | 23 | 110 | 102 | 81 | 73 | 25 | 105 | 55 | 17 | 68 | 93 | 113 |
|---|---|---|---|---|---|---|---|---|---|---|---|---|---|---|---|---|---|---|---|---|---|---|---|---|---|---|---|---|---|---|---|---|---|---|---|---|---|---|---|---|---|---|---|---|---|---|---|---|---|---|---|
| 15 | 15 | 15 | 15 | 15 | 15 | 15 | 16 | 16 | 16 | 16 | 16 | 16 | 16 | 16 | 16 | 16 | 16 | 16 | 16 | 16 | 17 | 17 | 17 | 17 | 17 | 17 | 17 | 17 | 17 | 17 | 18 | 18 | 18 | 18 | 18 | 18 | 18 | 18 | 18 | 18 | 18 | 18 | 18 | 18 | 18 | 19 | 19 | 19 | 19 | 19 | 19 |
| | | x | x | x | x | x | | x | x | | | | x | | x | x | | x | x | | x | x | x | | x | x | x | x | x | x | x | x | | x | x | x | x | x | x | | x | x | x | x | | | x | | x | x | |
| x | | | | | | | x | | | x | | | | x | | | | | | x | | | | x | | x | | | | x | | | x | | | | | | | x | x | | | | x | x | | | | | x |
| | x | | x | | x | x | | | x | | x | | | | x | x | x | | x | | | x | x | | | | x | x | | x | x | | x | | | x | | x | x | | x | x | x | x | | | x | | x | | |
| x | x | x | x | x | x | x | x | x | x | x | x | x | x | x | x | x | x | x | x | x | x | x | x | x | x | x | x | x | x | x | x | x | x | x | x | x | x | x | x | x | x | x | x | x | x | x | x | x | x | x | x |
| x | | | x | | | | | | | x | | x | | x | | | | | | x | | | | x | | x | | | | x | | | x | | | | x | | | x | | | | | x | | | | | | x |
| | | | | | | | | | | x | | | | | | | | | | | | | x | | | | | x | | | | | | | x | | | | | | | | | | | | | x | | | |
| x | x | x | x | x | x | x | | x | x | x | x | x | x | | x | x | x | x | x | | x | x | x | x | x | x | x | x | x | x | x | x | x | x | x | x | x | x | x | x | x | x | x | x | x | x | x | x | x | x | x |
| | | x | | x | | | | x | x | x | x | x | x | | | x | x | | | | | | | | x | x | x | | x | x | | x | | x | x | x | | x | x | x | | | x | x | x | x | x | x | x | x | x |
| | | x | | x | | | | | | x | | | x | | | x | x | | | | | | | | | | | | | | | | | | x | | | x | x | x | | | x | | x | | | | x | x | x |
| | | | | | | | x | | | x | | x | | | | | | | | x | | | | x | | x | | | | | | | x | | | | | | | | x | | | | | | | | | | x |
| x | | | | x | | | | | | x | | | | x | | x | | | | | | | | | x | | | | | | | | | x | x | | | | | x | | | x | | | | | | x | x | |
| | x | x | x | x | x | x | | x | x | x | | | x | | x | x | | x | x | | x | x | x | x | x | x | x | x | x | x | x | x | x | x | | x | x | x | x | x | x | x | x | x | x | x | x | | x | | x |
| | | x | x | x | | | | x | x | | | | | | | x | | | | | x | | | | x | x | | | x | x | | | | | | x | x | | | x | x | | | | x | x | | | | | x |
| | | x | x | | | | | x | | | | | x | | | | | | | x | x | | | x | | x | | | | x | | | | | | | x | | | x | | | | | | x | | x | | | x |
| | | x | x | | x | | | x | | x | | | x | | | | x | | | x | | | | | x | x | x | | x | x | | | | x | x | x | x | | | x | | | | | | | | | | | x |
| | | x | | | | | | | | | | | | | | | | | | x | | | | x | | | x | | x | | | | x | | x | | | | | x | | | | | | | | x | x | | x |
| | | x | x | | | | x | | | | x | | | x | x | | | | | x | | | | | x | | x | | x | | x | | | | x | | x | x | | | x | | x | x | | | | x | | | |
| x | | | x | x | | | | | | | | x | x | | | | | x | | | | | x | x | | x | x | | | | | | x | x | | x | x | | | x | x | x | x | x | x | | | | x | | x |
| | | | | x | | | | | | | | | | | | | | | | | | | x | | | | | | x | | | x | | x | | | | | | | | | x | x | | | | | x | | |
| | | x | | | | | | x | x | | | | x | | | x | x | | | | x | | | | x | | | | | | | | | | | | x | | | | | | | | | | | | | | |
| x | x | x | x | x | x | x | x | x | x | x | x | x | x | x | | x | x | x | x | x | x | x | x | x | x | x | x | x | | x | x | x | x | x | x | x | x | x | x | x | x | x | x | x | x | x | x | x | x | x | x |
| | | | | | | | | | | | | | | | x | | | | | | | | | | | | | | x | | | | | | | | | | | | | | | | | | | | | | |
| x | x | x | x | x | x | | x | x | | x | x | x | x | x | x | x | x | | x | x | x | x | | x | x | x | | x | x | x | x | x | x | x | | x | x | x | x | x | x | x | | x | x | x | | x | x | | x |
| | | x | | | | | | | | | x | | x | | | | | x | | | x | x | | x | x | x | x | | x | | x | | x | x | | x | x | x | x | x | x | | x | x | | x | | x | | | x |
| | | | | | | | | | | | | | | | | | | | | | | | | | | | | | | x | | | | | x | | | | | | | | | | | | x | | | x | |
| | | | | | | x | | | x | x | | | | | | | | | | | | | x | x | | | | | | | x | | | | | | | | | | | | x | | | | | | | | |
| x | x | | x | x | x | x | x | x | x | x | | x | | x | x | x | x | x | x | x | x | | x | x | x | | x | x | | x | x | | x | x | | | x | x | | x | x | x | x | x | x | x | | | x | | |
| | | x | | | | | | | | | x | | x | | | | | | | x | | x | | x | | x | | | x | x | | x | | | x | x | | | x | | | | | | | | x | x | | x | x |
| x | x | x | x | x | x | | x | x | x | x | x | x | x | x | x | x | x | x | | x | x | x | x | x | x | x | | x | x | | x | | x | x | | x | x | x | x | x | | x | x | x | x | x | | | x | | x |
| | | | | | | x | | x | | | | x | | | | | x | | x | | | x | x | x | | | x | x | x | | | | x | x | | x | x | x | | | x | x | | | x | x | | x | | | |
| | | | | | | | | | | | | | | | | | | | | | | | | | | | | | | x | x | x | | | x | | | | | | | | x | | | x | x | | | x | |
| | | x | | x | x | | x | x | x | | | x | x | x | x | x | x | | x | x | x | x | x | x | x | | | x | x | x | x | x | | x | x | x | | x | x | x | x | x | x | x | x | x | x | x | x | x | x |
| x | x | | x | | x | x | | | | x | x | | | | | | x | x | | | | | | | | x | x | | | | | | x | | | | x | | | | | | | | | | | | | | |
| x | x | x | | x | x | | | x | | | | x | | x | | | x | | x | x | | x | x | x | x | x | | x | | | x | x | | | x | x | x | | x | x | x | | | x | x | | x | | x | x | x |
| | | | | | | | | | | | | | | | x | | | x | | | | | | | x | | x | x | | x | x | x | | | | | x | x | x | | | | x | | | x | | | | | x |
| x | x | x | x | | | | | | | | x | x | | x | | | x | x | | x | | | | | | x | x | | | | | | | | | | | | | | | | x | | | | | | | | |
| | | | | | | | | | | | | | | x | | | | | | | | | | | | x | | | | | | | | | | | | | | | x | | | | x | | | | x | | |
| | | x | x | | | | | | x | | | | | | | | | | | | x | | | | | | | | | | | | | | | | | | | | | | | | | | | | | | |
| | | | | x | | x | | | | | | | | | x | | | | x | | | | x | | x | | | | | x | x | | | | | | | | x | x | x | x | x | x | x | | x | x | x | | x |
| | | | | | | | | | | | | | | | | | | | | | | | x | | | | | | | x | | | | | | | | | | | | | | | | | | | | | x |

| | | | | | | | | | | | | | | | | | | | | | | | | | | | | | | | | | | | |
|---|---|---|---|---|---|---|---|---|---|---|---|---|---|---|---|---|---|---|---|---|---|---|---|---|---|---|---|---|---|---|---|---|---|---|---|
| 83 | 45 | 43 | 91 | 133 | 62 | 19 | 33 | 114 | 92 | 98 | 94 | 28 | 97 | 29 | 82 | 89 | 53 | 134 | 88 | 54 | 57 | 63 | 135 | 64 | 75 | 119 | 115 | 77 | 136 | 79 | 95 | 67 | 78 | 20 | 65 |
| 19 | 19 | 19 | 19 | 19 | 19 | 19 | 20 | 20 | 20 | 20 | 20 | 20 | 20 | 20 | 20 | 20 | 20 | 20 | 20 | 20 | 20 | 20 | 20 | 20 | 20 | 20 | 20 | 20 | 20 | 20 | 20 | 20 | 20 | 20 | 20 |
| x | x | x | x | x | x | x | x |  | x |  | x | x | x |  | x |  | x | x | x | x | x | x | x | x |  |  |  | x | x | x | x | x |  | x | x |
|  |  |  |  |  |  |  |  | x |  | x |  |  |  | x |  | x |  |  |  |  |  |  |  |  | x | x | x |  |  |  |  |  | x |  |  |
| x | x |  |  | x | x | x |  |  | x | x |  |  | x |  | x |  | x | x | x | x | x |  |  | x |  | x |  | x |  | x | x |  |  | x | x |
| x | x | x | x |  | x | x | x | x | x | x | x | x | x | x | x | x | x | x | x | x | x | x |  | x |  | x |  | x | x | x | x | x | x | x | x |
|  | x |  |  |  |  |  | x | x |  |  |  |  |  |  |  | x |  |  |  |  |  |  |  |  | x |  | x |  |  |  |  |  | x |  | x |
|  |  | x |  |  |  |  | x |  | x |  |  |  |  |  |  | x |  |  |  |  |  | x | x |  |  | x |  |  |  |  | x |  |  |  | x |
| x | x | x | x | x | x | x | x | x | x | x | x | x | x | x | x | x | x | x | x | x | x | x | x | x | x | x | x | x | x | x | x | x | x | x | x |
|  |  | x | x |  |  | x |  | x |  |  |  | x | x | x |  | x |  | x |  | x |  |  |  |  |  |  |  |  | x | x |  | x | x |  | x |
|  |  | x |  |  |  |  | x | x |  |  |  |  | x |  |  | x |  | x |  | x |  |  |  |  |  |  |  |  |  |  |  |  | x |  |  |
|  |  |  |  |  |  |  |  |  |  |  |  |  |  |  |  | x |  |  |  |  |  |  |  |  | x |  |  |  |  |  |  |  |  |  |  |
|  |  |  | x |  |  | x | x | x |  |  |  |  |  |  |  |  |  |  | x |  |  |  |  |  |  |  |  |  |  |  |  |  | x |  |  |
| x | x | x | x | x | x | x | x | x | x |  | x | x |  | x | x | x | x | x |  | x | x |  | x | x |  |  | x | x | x | x | x | x | x | x | x |
|  |  |  |  |  |  |  |  | x |  |  |  |  |  |  |  | x |  |  |  |  |  |  |  |  |  |  | x |  | x |  |  |  | x |  |  |
|  | x | x | x |  | x |  |  | x |  |  |  |  |  | x |  | x |  |  |  |  |  |  |  |  |  |  | x |  | x |  |  |  | x |  |  |
|  |  | x | x | x | x |  |  | x |  |  |  | x | x | x | x | x |  |  |  |  |  |  |  |  |  |  | x |  |  |  |  | x | x |  |  |
| x |  |  |  | x |  |  | x |  | x |  |  |  |  |  |  |  | x | x | x | x |  |  |  |  | x | x |  |  |  |  | x |  |  | x | x |
| x |  |  | x |  |  |  |  | x |  | x | x |  |  | x |  | x |  |  |  |  | x |  | x |  |  |  |  |  |  |  |  |  |  |  |  |
|  |  |  |  |  |  |  |  | x |  |  |  | x | x | x | x |  |  |  | x |  | x |  | x |  | x | x | x |  |  | x | x | x |  |  | x |
| x |  |  |  |  |  |  |  | x |  |  |  |  |  |  |  |  |  |  |  |  | x |  | x |  |  |  |  | x |  |  |  |  |  |  |  |
|  |  |  |  |  | x | x |  | x |  |  |  |  |  | x |  |  |  |  |  |  |  |  |  |  |  |  |  |  |  |  |  |  | x |  |  |
| x | x | x | x | x | x | x | x | x | x | x | x | x | x | x | x | x | x | x | x | x | x | x | x | x | x | x | x | x | x | x | x | x | x | x | x |
|  |  |  |  |  |  |  |  |  |  |  |  |  |  |  |  |  |  |  |  |  |  |  |  |  |  |  |  |  |  |  |  |  |  |  |  |
| x |  | x | x | x |  | x | x | x | x | x | x |  | x | x | x | x |  | x | x |  | x |  | x |  | x | x |  |  | x |  | x | x | x |  | x |
| x |  | x | x |  |  |  |  | x | x |  |  |  | x | x | x |  |  |  |  |  | x |  | x |  |  | x |  | x | x |  |  | x | x |  |  |
|  |  |  |  |  | x |  |  |  |  |  |  | x |  | x |  |  |  |  |  | x |  | x |  | x |  |  | x |  | x |  | x |  |  | x | x |
|  | x |  |  |  |  |  | x |  |  |  |  |  |  | x | x |  | x |  |  |  |  |  |  |  |  |  |  |  |  | x |  | x |  |  |  |
|  | x |  | x | x |  | x | x |  | x | x | x |  | x | x | x | x | x | x | x |  | x |  |  |  | x | x |  | x | x | x | x |  | x |  | x |
| x |  | x |  |  | x |  |  | x | x |  |  | x |  | x |  |  |  |  |  | x |  | x | x | x |  |  | x |  | x |  | x | x |  | x |  |
| x | x |  |  | x |  | x | x | x |  | x |  |  | x | x | x | x | x | x | x |  |  |  | x |  | x | x |  |  |  | x |  | x |  |  |  |
|  |  | x | x | x |  |  | x | x | x | x |  |  | x | x |  |  |  |  |  |  | x |  |  |  |  |  |  |  |  |  |  |  |  |  |  |
|  |  | x |  |  | x |  | x | x | x |  | x | x |  | x |  |  | x |  |  | x |  | x |  | x |  |  | x | x | x |  | x |  | x | x | x |
| x | x | x | x | x | x | x | x | x | x | x | x | x | x | x | x | x | x | x |  | x | x | x | x | x |  | x | x | x | x | x | x | x | x | x |  |
|  |  |  |  |  |  |  |  |  |  |  |  |  |  |  |  |  |  |  | x |  |  |  |  |  | x |  |  |  |  |  |  |  |  |  | x |
| x | x | x | x | x |  |  | x | x |  |  | x |  | x |  | x | x | x | x | x | x | x |  |  | x | x | x | x | x | x |  | x | x | x | x | x |
| x | x |  | x |  |  |  |  | x |  |  |  |  |  |  |  |  |  |  |  |  |  |  |  |  |  |  |  | x |  |  |  |  |  | x | x |
|  |  |  |  |  |  |  |  | x |  |  |  |  |  |  |  |  |  |  |  |  |  |  |  |  | x |  |  |  |  |  |  |  |  |  | x |
|  |  |  |  |  |  |  |  |  |  |  |  | x |  |  |  |  |  |  |  |  |  |  |  |  |  |  |  |  |  |  |  |  |  |  |  |
|  |  |  |  |  |  |  |  |  |  |  |  |  |  |  |  |  |  |  |  |  |  |  |  |  |  |  |  |  |  |  |  |  |  |  |  |
| x |  |  |  |  | x |  |  | x | x | x |  | x |  | x | x | x | x |  |  | x |  |  |  | x |  | x | x | x |  | x |  | x | x |  |  |
|  |  |  |  |  |  |  |  |  |  |  |  |  |  |  |  | x |  |  |  |  |  |  |  |  |  | x |  |  |  |  |  |  |  |  |  |

**Table 2.** Public HAR Datasets Used in the Reviewed Literature (available as of 12/31/2020).

| Dataset | Environment | Population | Sensors | Sensor body location | Activities | Used in | Reference |
|---|---|---|---|---|---|---|---|
| WISDM* | Controlled | 36 | Accelerometer | Lower body part | Posture, mobility | [28,53–55,58,60,64,80–83,100,102,109–111] | [112] |
| UniMiB SHAR | Controlled | 30 | Accelerometer | Lower body part | Posture, mobility, other | [49,54,77,80,92] | [49] |
| Sussex-Huawei Locomotion (SHL) | Free-living | 3 | Accelerometer, gyroscope, magnetometer, other | Lower body part, upper body part, purse/backpack, hand | Mobility, locomotion | [23,78,113–115] | [23] |
| MobiAct | Controlled | 54 | Accelerometer, gyroscope, magnetometer | Lower body part | Mobility, other | [39,54,67,92] | [116] |
| Complex Human Activity** | Controlled | 10 | Accelerometer, gyroscope, magnetometer | Lower body part | Posture, mobility | [68,81,82,101] | [117] |
| Actitracker*** | Free-living | 225 | Accelerometer | NA / unconstrained | Posture, mobility | [64,104] | [118] |
| Extrasensory | Free-living | 60 | Accelerometer, gyroscope, magnetometer, GPS | NA / unconstrained | Posture, mobility, locomotion, other | [98,119] | [120] |
| Real World | Controlled | 15 | Accelerometer, gyroscope, magnetometer | Lower body part | Posture, mobility | [28,62] | [121] |
| Motion Sense | Controlled | 24 | Accelerometer, gyroscope | Lower body part | Posture, mobility | [28,92] | [122] |
| Sensor activity | Controlled | 10 | Accelerometer, gyroscope, magnetometer | Lower body part | Posture, mobility | [28] | [30] |
| Walking recognition | Controlled | 77 | Accelerometer, gyroscope, magnetometer | Lower body part, upper body part, purse/backpack, hand | Mobility | [29] | [29] |
| Real-life HAR | Free-living | 19 | Accelerometer, gyroscope, magnetometer, GPS | NA / unconstrained | Mobility, locomotion, other | [89] | [89] |
| Transportation Mode Detection | Free-living | 13 | Accelerometer, gyroscope, magnetometer, other | NA / unconstrained | Mobility, locomotion | [115] | [123] |
| HASC-2016 | Controlled, free-living | 567 | Accelerometer, gyroscope, magnetometer, other | Lower body part, upper body part, purse/backpack | Posture, mobility | [17] | [124] |
| Miao | Controlled | 7 | Accelerometer, gyroscope, magnetometer, other | Lower body part, upper body part | Mobility | [34] | [34] |

Note: NA = not available; * Also referred to as WISDM v1.1; ** Also referred to as Shoaib or SARD; *** Also referred to as WISDM v2.0.